%% file: main.tex
\documentclass{article}

\input{000-packages}
\input{001-commands}

\title{
Best of Both Worlds:
Robust Accented Speech Recognition \\
with Adversarial Transfer Learning
}
\input{002-authors}
\begin{document}
\ninept
\maketitle
\input{010-abstract}
\input{011-keywords}
\input{100-intro}
\input{200-related}
\input{300-approach}
\input{400-experiments}
\input{500-conclusion}

\bibliographystyle{IEEEbib}
\bibliography{refs}

\end{document}

%% file: 000-packages.tex
\usepackage{spconf}             % ICASSP required styling

\usepackage{amsmath}            % extensive math options
\usepackage{amssymb}            % special math symbols
\usepackage{bm}                 % bold math
\usepackage{bbm}                % math font
\usepackage{arydshln}           % for \cdashline, order seem to matter (put before booktabs?)
\usepackage{booktabs}           % for better tables
\usepackage{hyperref}           % for handling URLs
\usepackage{tabularx}           % more table options
\usepackage{graphicx}           % enhanced support for graphics
\usepackage{cite}               % automatically sort references
\usepackage{xcolor}             % more colour options
\usepackage{enumitem}           % more list formatting options
\usepackage{bbding}

\usepackage[disable]{todonotes}  % disable todo notes 

%% file: 001-commands.tex
\DeclareMathOperator*{\argmax}{argmax}
\DeclareMathOperator*{\argmin}{argmin}

%% file: 002-authors.tex
\name{
\begin{tabular}{c}
Nilaksh Das$^1$\sthanks{Work conducted as an intern at Amazon AWS AI.},
Sravan Bodapati$^2$,
Monica Sunkara$^2$,
Sundararajan Srinivasan$^2$,
Duen Horng Chau$^1$
\end{tabular}
}
\address{
$^1$Georgia Institute of Technology
\qquad
$^2$Amazon AWS AI
}

%% file: 010-abstract.tex
\begin{abstract}
Training deep neural networks
for automatic speech recognition (ASR) 
requires large amounts of transcribed speech.
This becomes a bottleneck for training robust models 
for \textit{accented} speech
which typically contains high variability
in pronunciation and other semantics,
since obtaining large amounts of annotated accented data 
is both tedious and costly.
Often, we only have access to large amounts 
of \textit{unannotated} speech from different accents.
In this work, we leverage this unannotated data 
to provide semantic regularization to an ASR model 
that has been trained only on one accent,
to improve its performance for multiple accents.
We propose Accent Pre-Training (Acc-PT),
a semi-supervised training strategy
that combines transfer learning and adversarial training.
Our approach improves the performance 
of a state-of-the-art ASR model
by 33\% on average over the baseline across multiple accents,
training only on annotated samples from one standard accent,
and as little as 105 minutes of \textit{unannotated} speech from a target accent.
\end{abstract}

%% file: 011-keywords.tex

%% file: 100-intro.tex
\vspace{-0.5em}
\section{Introduction}

Huge advancements in the field of 
automatic speech recognition (ASR)
have been made in the recent years 
by leveraging deep learning 
techniques~\cite{yu2016automatic,sainath2017multichannel,moritz2020streaming,kriman2020quartznet}.
State-of-the-art ASR systems nowadays
employ massive deep neural networks (DNNs),
which are trained on tremendous amounts of
annotated speech data.
This revolution in the ASR community
would not have been possible without
undertaking arduous data collection and annotation efforts
for human speech.
However, in practice,
there is usually limited diversity in the speech accent
for large annotated datasets that are used
to bootstrap ASR systems.
Moreover, datasets with higher multiplicity 
of annotated accented speech
are typically deficient in size, 
thus being rendered as inadequate for training
large ASR systems from scratch.
This degrades the generalization capacity of ASR systems
with respect to accented speech examples.
In this work, we aim to improve the accent robustness
of a trained state-of-the-art ASR model
in this low-resource setting for accent data.

While we have huge amounts of annotated data from certain locales
(e.g., the largest LibriSpeech dataset has primarily US-accent),
it is oftentimes the case in practice that
we only have access to \textit{unannotated} data from other locales
as it is easier to collect.
In this work, we focus on this setting to explore
how much gain can be made without undertaking the
expensive and tedious task of annotating accented speech.
Several semi-supervised learning techniques have been proposed
to tackle this scenario by combining annotated and unannotated examples
to jointly learn robust inference models\cite{synnaeve2019end,baskar2019semi,long2019large}.
We propose to leverage domain adversarial training (DAT),
one such semi-supervised approach~\cite{ganin2016domain} 
for our low-resource setting in the absence of annotated accented speech data.
DAT is an adversarial training technique 
that enforces intermediate representations to be domain-invariant
for different accented examples, 
and has been shown to improve
the accent robustness of ASR models with limited annotated data~\cite{sun2018domain}.
In this work, we augment this approach by proposing a training strategy that
incorporates transfer learning for further improving performance from DAT.

Transfer learning is a training paradigm that leverages a model trained
for a particular \textit{base} task, to reuse it as a starting point for 
training the model on another \textit{new} task.
This bootstrapping method is especially powerful 
when the new task has limited data, and is similar to the base task.
The bootstrapped model would have
some latent knowledge of the new task,
and does not require large amounts of new data.
In this paper, we leverage a state-of-the-art ASR model
that has been pre-trained for speech recognition for a base task,
and further train it with DAT for a new task in which we have no accented speech
annotations, and have limited accented data overall.
This mimics several practical scenarios in deploying real-world ASR systems.
Our work makes the following major contributions:
\begin{itemize}
    \item{
        We improve upon the semi-supervised DAT technique,
        combining it with transfer learning.
        Specifically, we propose to bootstrap 
        the discriminator 
        to accurately classify accented speech
        from the intermediate feature space of a trained encoder,
        before applying the DAT training.
        We call this approach ``Accent Pre-training'' (Acc-PT).
    }
    \item{
        We evaluate our approach across multiple accents, and show that our proposed Acc-PT technique
        combined with DAT improves the performance of the state-of-the-art ASR model
        by 33\% over the baseline across several accents on average, 
        despite having been trained only with annotated speech samples for one standard accent (US),
        and as little as 105 minutes of \textit{unannotated} speech from a target accent (Philippines).
    }
\end{itemize}

%% file: 200-related.tex
\vspace{-1em}
\section{Related Work}

Domain adversarial training (DAT) is a popular semi-supervised approach
used in several computer vision tasks~\cite{ganin2016domain,tzeng2017adversarial}.
It has been shown to improve the robustness of
ASR models in the low-resource setting, 
when limited annotated data is available~\cite{sun2018domain,tripathi2018adversarial,hu2019multilingual}.
DAT aims to make the intermediate latent space domain-invariant
by adversarially training a discriminator and encoder, 
thus regularizing the downstream decoder weight parameters
to be domain agnostic.
Learning such a domain-invariant acoustic feature space has also been shown
to improve ASR robustness in noisy environments~\cite{shinohara2016adversarial,serdyuk2016invariant}
and for speaker verification tasks~\cite{luu2020channel,meng2019adversarial,tu2019variational}.
For accent robustness, 
this method is leveraged to train the end-to-end ASR model (encoder and decoder)
with annotated data from a source accent, and simultaneously train
the encoder and discriminator combination with unannotated accented data
from target accents.
While previous work trains
the ASR model from scratch~\cite{sun2018domain}, we explore the added benefit
of leveraging transfer learning to improve a pre-trained ASR model
in the low-resource setting, where we have limited amount of
annotated source accent data and unannotated target accent data.
Furthermore, we propose the Accent Pre-training (Acc-PT) strategy
that bootstraps the discriminator by pre-training it on the 
learned representations from the model,
which is key to fully adapting DAT to work with transfer learning.

Several other approaches have been proposed recently to train accent robust DNNs for ASR.
However, many of these methods may fail in the low-resource setting.
A hierarchical grapheme-based technique is used to jointly learn 
grapheme and phoneme prediction tasks~\cite{rao2017multi}
using annotated data from multiple accents.
A multi-task learning (MTL) approach trains the ASR model jointly with
accented English and native language annotated speech 
from Spanish and Indian speakers~\cite{ghorbani2018leveraging}.
Some MTL methods for improving accent robustness also take an inverse approach than DAT by 
increasing the domain variance of the intermediate latent representation,
thus jointly learning an ``accent aware'' ASR model~\cite{jain2018improved, viglino2019end}.
However, to jointly train such MTL models, 
annotated samples are required from \textit{multiple} accents,
posing a stark contrast to the DAT approach which requires
annotated samples from only the source accent.
For example, MTL models~\cite{jain2018improved, viglino2019end}
train on annotated speech samples from 7 accents, 
whereas our DAT approach only trains on annotated speech samples
from just one accent.
This demonstrates the versatility of the DAT approach, especially considering
the low-resource setting.
Previous DAT work outlines this distinction with MTL,
showing that DAT outperforms MTL with limited data resources~\cite{sun2018domain}.
Hence, in this paper,
we specifically focus on further improving the DAT approach 
by combining it with transfer learning.

%% file: 300-approach.tex
\section{Approach}

\begin{figure}
    \centering
    \includegraphics[width=\linewidth]{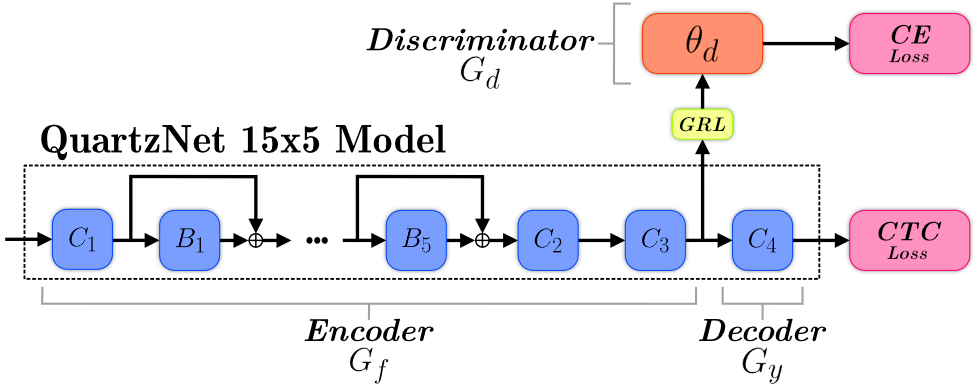}
    \caption{DAT with QuartzNet 15x5 model.}
    \vspace{-1.5em}
    \label{fig:dat_arch}
\end{figure}

Our main idea is to leverage 
the annotated speech of a standard accent, 
which is abundantly available in general,
to improve the robustness of a trained state-of-the-art ASR model
on the unannotated speech of other unseen accents.
In this work, we denote the standard accent data 
having seen annotations as 
$S = \{\bm{x}_i, \bm{y}_i, a_i=0\}_{i=1}^{|S|}$
and the other accented speech samples with different accents 
having unseen annotations as
$U = \{\bm{x}_i, a_i \geq 1\}_{i=1}^{|U|}$. 
In this notation, $\bm{x}_i$ and $\bm{y}_i$
are the input speech spectrogram features and the respective transcriptions,
and $a_i$ is the corresponding accent for a given sample.

\input{301-conventional}
\input{302-dat}
\input{303-acc_pt}

%% file: 301-conventional.tex
\vspace{-1em}
\subsection{End-to-end Speech Transcription with CTC loss}

\input{3xx-quartznet_layers}

For our baseline, 
we use a trained QuartzNet 15x5 model~\cite{kriman2020quartznet},
which is a state-of-the-art end-to-end neural acoustic model for ASR,
and is made publicly available\footnote{\url{https://ngc.nvidia.com/catalog/models/nvidia:quartznet_15x5_ls_sp}} 
by the authors.
QuartzNet is based on the Jasper architecture~\cite{li2019jasper},
a fully convolutional (conv) model 
trained with Connectionist Temporal Classification (CTC) loss~\cite{graves2006connectionist}.
A representation of the model architecture can be seen in Figure~\ref{fig:dat_arch}.

The first layer of the QuartzNet 15x5 model is a 1D conv layer $C_1$ 
having a stride of 2,
which is followed by a sequence of blocks. 
Each block $B_i$ is repeated 3 times,
and has residual connections between blocks. 
Further, a block $B_i$ consists of the following group of 4 layers, 
the group being repeated 5 times within the block:
1) $K$-sized depthwise conv layer with $c_{\text{out}}$ channels;
2) a pointwise conv operation;
3) a batch-wise normalization layer;
and 4) a ReLU activation.
The final stage of the QuartzNet model has
3 more conv blocks $(C_2, C_3, C_4)$,
where $C_4$ (having a dilation of 2)
does the final decoding of labels.
Table~\ref{tab:quartznet_layers} gives more detailed architectural specifications of the blocks.

In this work, we use the QuartzNet model that has been trained 
on the 960-hour LibriSpeech train set~\cite{panayotov2015librispeech},
achieving a near state-of-the-art ASR performance
with WER=3.90\% on the ``test-clean'' split
and WER=11.28\% on the ``test-other'' split,
without using any language model. 
The model was trained for 400 epochs
using the NovoGrad optimizer~\cite{ginsburg2019stochastic}
with data augmentation including
speed perturbation combined with Cutout~\cite{devries2017improved}.

Using a pre-trained QuartzNet model allows us to
establish a transfer learning baseline,
where ASR performance on LibriSpeech 
(a dataset with limited accent diversity) 
constitutes the base task.
Our new task for transfer learning
is to perform robust transcription
on a dataset with multiple accents
(e.g., Common Voice dataset, described in Section \ref{sec:data}).
Our main idea is to leverage the latent knowledge
learned by the trained model for recognizing human speech
(not necessarily for different accents),
and augment its generalization capacity
by re-tuning it in a semi-supervised fashion
in the absence of annotated accented data.

%% file: 3xx-quartznet_layers.tex
% \begin{table}[t]
% \begin{center}
% \resizebox{\linewidth}{!}{%
% \begin{tabular}{l|ccccccccc}
% \toprule
% % & \multicolumn{8}{c|}{\textit{Encoder}} & \textit{Decoder} \\
% Block & $C_1$ & $B_1$ & $B_2$ & $B_3$ & $B_4$ & $B_5$ & $C_2$ & $C_3$ & $C_4$ \\
% \midrule
% $K$ & 33 & 33 & 39 & 51 & 63 & 75 & 87 & 1 & 1\\
% $c_{\text{out}}$ & 256 & 256 & 256 & 512 & 512 & 512 & 512 & 1024 & $\|$labels$\|$\\
% \bottomrule
% \end{tabular}
% }
% \caption{The QuartzNet 15x5 architecture.}
% \vspace{-1em}
% \label{tab:quartznet_layers}
% \end{center}
% \end{table}

\begin{table}[t]
\begin{center}
\begin{tabular}{l|cc}
\toprule
Block & $K$ & $c_{\text{out}}$ \\
\midrule
$C_1$ & 33 & 256 \\
$B_1$ & 33 & 256 \\
$B_2$ & 39 & 256 \\
$B_3$ & 51 & 512 \\
$B_4$ & 63 & 512 \\
$B_5$ & 75 & 512 \\
$C_2$ & 87 & 512 \\
$C_3$ & 1  & 1024 \\
$C_4$ & 1  & $\|$labels$\|$ \\
\bottomrule
\end{tabular}
\caption{The QuartzNet 15x5 architecture.}
\vspace{-2.5em}
\label{tab:quartznet_layers}
\end{center}
\end{table}

%% file: 302-dat.tex
\subsection{Domain Adversarial Training (DAT)}

DAT is an adversarial training approach 
that aims to learn an intermediate latent feature space
that is domain-invariant.
A DAT model consists of 3 main components:
1) the feature encoder network $G_f$ with parameters $\theta_f$, 
such that $\bm{f} = G_f(\bm{x}; \theta_f)$, 
where $\bm{x}$ are the input speech features;
2) the domain classifier (discriminator) network $G_d$ with parameters $\theta_d$,
such that $a = G_d(\bm{f}; \theta_d)$,
where $a$ is the class label, which is the accent in our case;
and 3) the task decoder network $G_y$ with parameters $\theta_y$,
such that $\bm{y} = G_y(\bm{f}; \theta_y)$,
where $\bm{y}$ is the inferred transcription for an ASR model.
Figure~\ref{fig:dat_arch} shows this setup.
The goal of DAT is to make the feature $f$
generated by $G_f$ to be domain-invariant
(i.e., accent-invariant for our ASR task)
for inputs from any accent $\bm{x} \in S$ or $\bm{x} \in U$,
so that the decoder network $G_y$ can be domain-agnostic
and robust to unseen accents.
For our QuartzNet model, we pick the intermediate layer $C_3$
for generating domain-invariant features.

The DAT objective function for a batch of $N$ samples is written as follows:
\begin{equation}
    \label{eq:datloss}
    E(\theta_f, \theta_y, \theta_d) = \\
        \frac{1}{N} \sum_{i=1}^N 
            % \big(
                \mathbbm{1}_{a_i=0} \mathcal{L}_y^i(\theta_f, \theta_y)
            % \big)
            - \lambda \mathcal{L}_d^i(\theta_f, \theta_d)
\end{equation}

\noindent
For our ASR task with the QuartzNet model, 
$\mathcal{L}_y^i(\theta_f, \theta_y)$ is the CTC loss 
for decoding transcription characters,
and $\mathcal{L}_d^i(\theta_f, \theta_d)$ is the cross-entropy (CE) loss
for accent classification. Note that $\mathcal{L}_y^i(\theta_f, \theta_y)$
also has an indicator function as the coefficient, 
ensuring that CTC loss is only present for the annotated samples
from the seen accent.

To optimize the parameters, we find the saddle points 
for the objective function as:
\begin{align}
    (\hat{\theta}_f, \hat{\theta}_y) &= 
        \argmin_{\theta_f, \theta_y} E(\theta_f, \theta_y, \hat{\theta}_d) \\
    \hat{\theta}_d &= 
        \argmax_{\theta_d} E(\hat{\theta}_f, \hat{\theta}_y, \theta_d)
\end{align}

\noindent
In DAT, this ``min-max'' optimization is done simultaneously
within a single backward pass by using the gradient reversal layer (GRL)
between feature encoder $G_f$ and domain classifier $G_d$.
The GRL layer acts as the identity function in the forward pass,
but in the backward pass it multiplies the incoming gradients with $-\lambda$.
Hence, the SGD update equations for DAT with a learning rate of $\mu$ can be written as follows:
\begin{align}
    \theta_f &\longleftarrow 
        \theta_f - \mu \frac{1}{N} \sum_{i=1}^N  
        \Bigg(
            \mathbbm{1}_{a_i=0}
            \frac{\partial \mathcal{L}_y^i}{\partial  \theta_f}
            - \lambda \frac{\partial \mathcal{L}_d^i}{\partial  \theta_f}
        \Bigg) \\
    \theta_y &\longleftarrow 
        \theta_y - \mu \frac{1}{N} \sum_{i=1}^N 
        \mathbbm{1}_{a_i=0}
        \frac{\partial \mathcal{L}_y^i}{\partial  \theta_y} \\
    \theta_d &\longleftarrow 
        \theta_d - \mu \frac{1}{N} \sum_{i=1}^N \lambda \frac{\partial \mathcal{L}_d^i}{\partial  \theta_d}
\end{align}

\noindent
Note here that $\lambda > 0$ corresponds to 
adversarially training the model with DAT, 
while $\lambda = 0$ will simply ignore the unannotated accent samples,
thus only performing conventional CTC training.
% Further, $\lambda < 0$ simply corresponds to the MTL setting.

%% file: 303-acc_pt.tex
\subsection{Accent Pre-training for Robust DAT}

\input{3xx-init_weights}

In the DAT approach, the feature encoder and discriminator networks
have competing objectives. 
Hence, during training, the feature encoder network 
is continuously learning from the gradients of the discriminator
so as to generate domain-invariant features that would ``fool'' the discriminator.
Nevertheless, if we start training with an untrained discriminator,
it will send highly noisy signals back to the encoder, 
resulting in unstable training.
A strategy was proposed to mitigate this by initializing $\lambda = 0$
and gradually increasing its value over the epochs~\cite{ganin2016domain}.
We integrate this strategy in our experiments.
However, we observe that when performing DAT combined with transfer learning,
such fundamental parameter update strategies are not sufficient 
as we start with an already powerful encoder network.
In this scenario, the discriminator is never able to ``catch up''
with the encoder, thus degrading the effect of adversarial training.
This issue is further exacerbated for transfer learning 
in the low-resource setting,
when there is insufficient data to train 
a randomly initialized discriminator jointly
with a fully tuned encoder.

In this work, we propose Accent Pre-training (Acc-PT) to overcome this issue, 
making DAT robust for low-resource transfer learning.
We hypothesize that performing DAT transfer learning 
with a strong discriminator trained on the initial latent space 
leads to more stable training, 
since both encoder and discriminator are at an equal footing.
In our Acc-PT approach, we initially freeze the feature encoder network
and pre-train the discriminator until convergence.
Once we have a pre-trained discriminator for the pre-trained ASR model,
we then perform domain adversarial training.
Table~\ref{tab:weights} outlines the pre-trained parameters for different training regimes.
In our experiments, we empirically show that Acc-PT combined with DAT
improves upon the performance of DAT for multiple accents 
when we re-tune the QuartzNet model.

%% file: 3xx-init_weights.tex
\begin{table}[b]
\begin{center}
\begin{tabular}{l|ccc}
\toprule
& $\theta_f$ & $\theta_y$ & $\theta_d$ \\
\midrule
Transfer Learning & yes & yes & - \\
% \rotatebox[origin=c]{180}{$\Lsh$} 
\hspace{0.2cm}
+ DAT & yes & yes & no \\
% \rotatebox[origin=c]{180}{$\Lsh$} 
\hspace{0.2cm}
+ Acc-PT + DAT \textbf{(proposed)} & yes & yes & \textbf{yes} \\
\bottomrule
\end{tabular}
\caption{Pre-trained parameters, ``-'' = not used.}
\label{tab:weights}
\end{center}
\end{table}

%% file: 400-experiments.tex
\section{Experiments}

\input{4xx-data}
\input{4xx-wers}
\input{401-data}
\input{402-baseline}
\input{403-results}

%% file: 4xx-data.tex
\begin{table}[b]
\begin{center}
\begin{tabular}{lrr}
\toprule
    Accent & \# utterances & \# hours \\
\midrule
US          & 206,653 & 255.3 \\
\cdashline{1-3}[.7pt/1.5pt]
England     & 76,622 & 91.3 \\
Indian      & 31,919 & 42.0 \\
Australia   & 29,521 & 38.0 \\
Scotland    & 8,405 & 11.9 \\
African     & 5,430 & 7.0 \\
Philippines & 1,895 & 2.5 \\
\midrule
Total       & 360,445 & 448 \\
\bottomrule
\end{tabular}
\caption{Details of different accents extracted from the validated samples of the Mozilla Common Voice dataset.}
\label{tab:data}
\end{center}
\end{table}

%% file: 4xx-wers.tex
\begin{table*}[t]
\begin{center}
\begin{tabular}{l|r|rrrrrr|rr}
\toprule
         &    &         \multicolumn{6}{c|}{\textit{Unseen accents for CTC during training}}    & Wt. Avg. & Wt. Avg. \\
         & US & England & Indian & Australia & Scotland & African & Philippines & \textit{unseen} & \textit{all} \\
         & (41,330) & (15,321) & (6,384) & (5,904) & (1,681) & (1,087) & (379) & (30,756) & (72,086) \\
\midrule
Baseline & 15.64 & 17.81 & 38.46 & 26.09 & 51.15 & 19.20 & 27.86 & 25.68 & 19.92 \\
% \rotatebox[origin=c]{180}{$\Lsh$}
\hspace{0.2cm}
+ DAT & 8.88 & 14.62 & 26.29 & 21.74 & 42.80 & 13.99 & 18.62 & 19.98 & 13.61 \\
% \rotatebox[origin=c]{180}{$\Lsh$} 
\hspace{0.2cm}
+ Acc-PT + DAT & 8.92 & 14.77 & 26.26 & 21.99 & 43.10 & 14.28 & 18.23 & 20.11 & 13.70 \\
Conventional CTC & 8.98 & 13.88 & 25.96 & 21.20 & 42.14 & 13.92 & 18.31 & 19.39 & 13.42 \\
% \rotatebox[origin=c]{180}{$\Lsh$} 
\hspace{0.2cm}
+ DAT & 8.84 & 14.03 & \textbf{25.59} & 21.30 & 42.32 & 13.87 & 18.42 & 19.42 & 13.35 \\
% \rotatebox[origin=c]{180}{$\Lsh$} 
\hspace{0.2cm}
+ Acc-PT + DAT & \textbf{8.81} & \textbf{14.01} & 25.64 & \textbf{20.71} & \textbf{42.08} & \textbf{13.85} & \textbf{18.09} & \textbf{19.29} & \textbf{13.28} \\
\bottomrule
\end{tabular}
\caption{Word error rates (WERs in \%) for different approaches computed across multiple accents. Annotated speech from only the US accent is used while training with CTC, all other accents are treated as unannotated in our experiments. The numbers in parenthesis denote the number of utterances for each accent in the test set.}
\vspace{-1.5em}
\label{tab:wers}
\end{center}
\end{table*}

%% file: 401-data.tex
\subsection{Data}
\label{sec:data}

We use Mozilla's Common Voice dataset~\cite{mozillacommonvoice}
for our experiments. 
Common Voice is a crowd-sourced project that aims to
collect natural human speech in different languages
from a variety of demographics of people across the world.
The source text for the speech is collected from various
sources such as blog posts, movie dialogues etc.
Since our QuartzNet model is pre-trained to recognize English speech,
we use the English subset from Common Voice,
which was reported to have 1.7k hours of speech at the time of access,
with 1.3k hours of speech having been validated for correctness
by majority voting.
Contributors to this dataset can also self-report their speech accent
which gets annotated for each speech sample.
This is of most use to us since this allows us to extract
accented speech with accent labels from the dataset.
The source text is repeated multiple times across and within accents,
making this dataset a good candidate for extracting accent-specific semantics.

We picked US accent as the standard seen accent 
as it has the largest amount of annotated speech ($\sim$255 hours).
We chose 6 other accents as the target unseen accents,
for which we only use the speech and ignore the annotations.
We chose 3 accents (England, Indian, Australia) having
relatively higher availability,
and 3 accents (Scotland, African, Philippines) having
very low availability in the dataset.
For all 7 accents, we only use the validated speech samples,
which total up to 448 hours.
Note that all 7 accents combined is still less than
half of 960 hours of LibriSpeech data used to train
the QuartzNet model.
Table~\ref{tab:data} shows a summary of the all the accent data.
For train, test and validation splits, we randomly partition
the accented samples into 70\%, 20\% and 10\% respectively.

%% file: 402-baseline.tex
\subsection{Baseline}

All ASR models in this work were trained using the 
NVIDIA NeMo toolkit~\cite{nvidianemo}.
We use the QuartzNet 15x5 model~\cite{kriman2020quartznet} that was
pre-trained on LibriSpeech for our transfer learning baseline.
Without any re-training, the baseline model has a micro average word error rate (WER) of 19.92\%
across all accents (Table~\ref{tab:wers}).
For DAT, we treat the output of block $C_3$ from the QuartzNet 15x5 model
as the domain-invariant latent space. 
Hence, the QuartzNet blocks $\{C_1, B_1, \dots C_2, C_3\}$ can be seen
as our feature encoder, and block $C_4$ as the task decoder.
For the DAT discriminator input, 
we take the mean of the encoded sequence embeddings
from the output of block $C_3$
as the generated feature $\bm{f}$.
The discriminator architecture consists of 
a linear layer of output size 512
followed by 2 dense layers of output size 1024,
finally ending with a linear layer which outputs 
the class scores with softmax acivation.
Each dense layer consists of 
a linear layer, ReLu activation and dropout.
For all re-training and DAT, we used the NovoGrad optimizer,
which was used to originally train our baseline model.
The network hyperparameters (e.g., $\lambda$) were tuned for best overall performance
on the validation split.

%% file: 403-results.tex
\subsection{Results}

We summarize all results from our experiments in Table~\ref{tab:wers}.
The last 2 columns show the weighted average WER performance
for the unseen accents
as well as all accents combined.

\medskip
\noindent
\textbf{Accent-invariant transcription.}
We see that even simple DAT with re-tuning the baseline model
significantly boosts the performance across all accents.
This can be attributed to the fact that there is a domain mismatch between the 
base task (LibriSpeech; more formal speech from audio books) 
and the new task (Common Voice; more informal speech from various sources such as movie dialogues).
However, due to the limited amount of accented data available in the Common Voice dataset,
Acc-PT does not improve upon the DAT performance from the baseline model
even though we observed a high discriminator accuracy.
More specifically, due to the domain mismatch between the base task and new task,
the DAT training with an accent discriminator pre-trained on the base task
introduces noise corresponding to the base task for accent regularization,
thus driving the latent space away from the optimal target distribution for the new task.

\medskip
\noindent
\textbf{Leveraging transfer learning for domain shift.}
We perform conventional CTC transfer learning by re-tuning
the baseline model on the standard accent (US-English). 
We observe that it improves upon DAT performance since 
there is no longer any incoming signals from the discriminator 
that is domain specific to the base task.
Comparing transfer learning performance of DAT and Acc-PT+DAT on the
conventional CTC model reveals that Acc-PT is now able to
better regularize the latent space to make it accent invariant
with respect to the optimal target distribution for the new task.
Our Acc-PT+DAT approach shows the best WER across 5 out of 6
unseen accents.

Additionally, we observe that simple DAT in this case improves overall WER (from 13.42\% to 13.35\%)
mainly due to improvement in US accent, whereas the performance on
unseen accents actually degrades with respect to conventional CTC (from 19.39\% to 19.42\%).
The noise backpropagated from an untrained, randomly initialized discriminator
during the initial epochs leads to sub-optimal adversarial training.
In contrast, Acc-PT with a pre-trained discriminator leads to
better DAT training with transfer learning, 
even improving the US-accent performance
along with having the best unseen accent performance.

We also note here that the magnitude of improvement 
of our Acc-PT+DAT approach over the other methods (e.g., from 19.39\% to 19.29\%)
could be much more pronounced for ASR models 
that are already not as powerful as the QuartzNet 15x5 model that we use in this study.
We intentionally choose a highly robust baseline so that 
our results are beneficial for practitioners
who are interested in deploying real-world ASR systems.

%% file: 500-conclusion.tex
\vspace{-1em}
\section{Conclusion}

In this work, we explore domain adversarial training (DAT) combined with transfer learning
to improve the accent robustness of a trained ASR model in the low-resource setting.
We propose the Accent Pre-training (Acc-PT) strategy of pre-training the discriminator 
on unannotated accented speech samples when performing DAT with a trained model.
Our experiments show that Acc-PT with DAT improves the performance
of the ASR model by 33\% on average across multiple accents,
when annotated samples from only a single accent were available.
In future work, we aim to develop visualization techniques
for interpreting the latent space with respect to different accents,
which would guide architectural decisions, choice of hyperparameters,
and model debugging while performing DAT for accent robustness.
This would also allow us to qualitatively compare DAT
with other end-to-end accent robustness approaches.